\documentclass[amsmath, amssymb, preprintnumbers, showpacs,
 showkeys,aps,prl,superscriptaddress,twocolumn]{revtex4-1}

\usepackage{color} 
\usepackage{hyperref}
\usepackage{amsmath,amssymb,mathrsfs}
\usepackage{psfrag}
\usepackage{graphicx}
\usepackage{graphics}
\usepackage{subfigure}
\usepackage{epsfig}
\usepackage{bm}
\usepackage{verbatim,color,ulem}
\usepackage{verbatim}
\usepackage{soul}

\begin{document}
	
\title{Collisionless Dynamics in Two-Dimensional Bosonic Gases}
	 
\author{A. Cappellaro} 
\affiliation{Dipartimento di Fisica e Astronomia ``Galileo Galilei'', 
Universit\`a di Padova, via Marzolo 8, 35131 Padova, Italy} 
\author{F. Toigo} 
\affiliation{Dipartimento di Fisica e Astronomia ``Galileo Galilei'', 
Universit\`a di Padova, via Marzolo 8, 35131 Padova, Italy} 
\author{L. Salasnich} 
\affiliation{Dipartimento di Fisica e Astronomia ``Galileo Galilei'', 
Universit\`a di Padova, via Marzolo 8, 35131 Padova, Italy} 
\affiliation{CNR-INO, via Nello Carrara, 1 - 50019 Sesto Fiorentino, Italy}

\begin{abstract}
We study the dynamics of dilute and ultracold bosonic gases 
in a quasi two-dimensional (2D) configuration and in the collisionless regime. 
We adopt the 2D Landau-Vlasov equation to describe a three-dimensional gas 
under very strong harmonic confinement along one direction. 
We use this effective equation to investigate 
the speed of sound in quasi 2D bosonic gases, 
i.e. the sound propagation around a Bose-Einstein 
distribution in collisionless 2D gases. We derive coupled 
algebraic equations for the real and imaginary parts of the sound velocity,  
which are then solved taking also into account the equation 
of state of the 2D bosonic system. Above 
the Berezinskii-Kosterlitz-Thouless critical temperature 
we find that there is rapid growth of the imaginary 
component of the sound velocity which implies a strong Landau damping. 
Quite remarkably, our theoretical results are in good agreement with very 
recent experimental data obtained with a 
uniform 2D Bose gas of $^{87}$Rb atoms.  
\end{abstract}

\date{\today}

\maketitle 

\textit{Introduction}. The Boltzmann-Vlasov equation is the most 
relevant tool to investigate the kinetics of three-dimensional (3D) 
quantum gases made of out-of-condensate atoms \cite{kirkpatrick1983,
kirkpatrick1985-lowtemp,zaremba1998,nikuni1999,zaremba1999-lowtemp,
griffin-book}. In the collisionless regime this equation reduces to the 
Landau-Vlasov equation, where the collisional integral is neglected 
but the mean-field interaction potential is still present and 
supports collective modes \cite{vlasov1945,landau1946,landau}. 
In the case of fermionic gases the speed of sound in this 
collisionless regime is the well-know zero-sound velocity 
of fermions around the Fermi-Dirac distribution \cite{landau,baym-book}. 

In two-dimensional (2D) uniform systems the Mermin-Wagner-Hohenberg 
theorem \cite{mermin,hohenberg} precludes Bose-Einstein condensation 
at finite temperature, but quasi-condensation and superfluidity 
is possible below the Berezinskii-Kosterlitz-Thouless critical 
temperature $T_c$ \cite{berezinskii,kosterlitz}. Very recently 
the speed of sound in a uniform quasi-2D Bose gas made of $^{87}$Rb atoms 
has been measured \cite{dalibard2017-pra,dalibard2017-ictp}. 
These experimental results are in agreement with 
theoretical predictions \cite{ota2017} based on the two fluids
hydrodynamics of Landau-Khalatnikov only  { well below $T_c$}. 

{ The authors of \cite{dalibard2017-pra,dalibard2017-ictp} explain 
the discrepancy above $T_c$ by suggesting that the experimental conditions 
are such that in this case collisions are not efficient enough 
to ensure the local thermodynamic equilibrium required by hydrodynamics 
and therefore the dynamics is collisionless.} 

In this paper { we suppose that also below $T_c$, 
where the superfluid component is present,  
the dynamics of the normal component is collisionless. 
and therefore the dynamics of the whole fluid is not collisional.} 
To substantiate this hypothesis we 
investigate the collisionless regime { by} 
using an effective 2D Landau-Vlasov equation. 
We study the speed of sound around a spatially-uniform 
Bose-Einstein distribution. We derive algebraic formulas 
for the real and imaginary parts 
of the speed of sound as a function of both temperature and interaction 
strength. Quite remarkably, our theoretical results for the 
real part of the sound velocity are in good agreement 
with the experimental data of Ref. 
\cite{dalibard2017-pra,dalibard2017-ictp}. Moreover, we find that 
the imaginary part of the sound velocity is negligible below 
the critical temperature $T_c$ while it becomes sizable  
close and above $T_c$, again in agreement with the recent 
experiment \cite{dalibard2017-ictp}. 

\textit{Kinetic approach for the 2D Bose gas}. 
Let us begin by considering a dilute and ultracold three-dimensional (3D) 
gas made of $N$ identical bosonic atoms of mass $m$, whose mutual interaction 
is modelled through a zero-range pseudo-potential where 
$g= 4\pi\hbar^2 a_s/m$ is the 3D interaction strength and 
$a_s$ the 3D s-wave scattering length. We assume that the bosonic system 
is under external confinement given by the trapping potential 
\begin{equation}
U_{\text{ext}}(\mathbf{r},z)= \mathcal{U}(\mathbf{r}) 
+ \frac{1}{2}m\omega_z^2 z^2 \; ,   
\end{equation}
that is the sum of a generic potential $\mathcal{U}(\mathbf{r})$ 
in the plane $x-y$ with $\mathbf{r}=(x,y)$ the 2D position
and a harmonic confinement along the $z$ axis. 

An effective two-dimensional (2D) configuration can be realized when
the harmonic confinement along the $z$ axis is tigt enough. 
In order to effectively constrain atoms on a plane, 
the energy $\hbar \omega_z$ of longitudinal confinement must be much
larger than the planar average kinetic energy $(p_x^2+p_y^2)/(2m)$ 
with $\mathbf{p}=(p_x,p_y)$ the planar linear momentum, 
a condition actual experiments can provide quite easily. 
The 3D system is then forced to occupy the longitudinal 
ground state along the confining axis and one finds \cite{baldovin2016} 
that the planar distribution  $f(\mathbf{r},\mathbf{p})$ 
of atoms in the 4D single-particle phase space ($(\mathbf{r,p})=(x,y,p_x,p_y)$)
satisfies the effective 2D Landau-Vlasov equation \cite{landau,baym-book}. 
\begin{equation}
\bigg[\frac{\partial}{\partial t} +  
\frac{\mathbf{p}}{m}\cdot \nabla_{\mathbf{r}}
- \nabla_{\mathbf{r}} \Big(\mathcal{U} + \mathcal{U}_{\text{mf}} 
\Big)\cdot \nabla_{\mathbf{p}} \bigg]f(\mathbf{r},\mathbf{p},t) = 0 \; , 
\label{2d boltzmann-vlasov}
\end{equation}
where 
\begin{equation}
\mathcal{U}_{\text{mf}}(\mathbf{r},t) = g_{2D} 
\int \frac{d^2\mathbf{p}}{(2\pi\hbar)^2} \; f(\mathbf{r},\mathbf{p},t) \; 
\end{equation} 
is the self-consistent Hartree-Fock dynamical 
mean-field term \cite{griffin-book,prokofev2001,prokofev2002}, 
and the memory of the original 3D character of the system 
is encoded in the renormalized 2D coupling constant 
\begin{equation}
g_{2D}=\frac{\sqrt{8\pi}\hbar^2}{m}\bigg(\frac{a_s}{a_z}\bigg)
\label{rescaled 2D g}
\end{equation}
with $a_z = \sqrt{\hbar/(m\omega_z)}$ the characteristic length 
of the axial harmonic confinement. 

\textit{Collective dynamics in collisionless 2D Bose gas}.  
The calculation of transport quantities 
requires the solution of Eq. \eqref{2d boltzmann-vlasov}. 
In the following we prove that 
a collisionless dynamical description based 
on Eq. \eqref{2d boltzmann-vlasov} recovers experimental data obtained in 
a homogeneous configuration of area $L^2$, realized by implementing 
a box potential on the $x-y$ plane \cite{dalibard2017-pra,
dalibard2017-ictp}. Thus, we set $\mathcal{U}(\mathbf{r})=0$ and also 
\begin{equation}
f(\mathbf{r},\mathbf{p},t) = f_0({\bf p}) + 
\delta f(\mathbf{r},\mathbf{p},t) 
\end{equation}
where $f_0(\mathbf{p})$ is a stationary and isotropic distribution 
and $\delta f(\mathbf{r},\mathbf{p},t)$ 
a very small perturbation around it. It follows that 
the linearized Landau-Vlasov 
equation for $\delta f(\mathbf{r},\mathbf{p},t)$ reads 
\begin{widetext}
\begin{equation}
\bigg[\frac{\partial}{\partial t} 
+ \frac{\mathbf{p}}{m}\cdot \nabla_{\mathbf{r}} 
\Bigg] \delta f(\mathbf{r},\mathbf{p},t) 
= g_{2D} \int {d^2\mathbf{p}'\over (2\pi\hbar)^2} \nabla_{\mathbf{r}} 
\delta f(\mathbf{r},\mathbf{p}',t) \cdot
\nabla_{\mathbf{p}} f_0(\mathbf{p}) \; .
\label{linearized 2d boltzmann-vlasov eq}
\end{equation}
\end{widetext}
Performing the Fourier transform of this equation 
according to 
${\widehat{\delta f}}(\mathbf{k},\mathbf{p},\omega)= 
\int {dt} \int {d^2\mathbf {r}} \, {\delta f}(\mathbf{r},\mathbf{p},t)  
\exp{\left(i (\mathbf{k} \cdot \mathbf{r}- \omega t)\right)}$ 
with $\mathbf{k}$ a 2D wavevector and $\omega$ the angular 
frequency, one finds an implicit formula for the dispersion relation 
\cite{landau}, given by
\begin{equation}
1 - g_{2D} \int {d^2\mathbf{p}\over (2\pi \hbar)^2} \;
\frac{\mathbf{k} \cdot \nabla_{\mathbf{p}}f_0(\mathbf{p})} 
{\mathbf{p}\cdot\mathbf{k}/m-\omega}= 0\;.
\label{susceptibility}
\end{equation} 
Note that this equation is nothing else than the condition to 
find the pole of the dynamic response function of the system 
within the random-phase approximation (RPA) \cite{baym-book}. 
{ Equation (\ref{linearized 2d boltzmann-vlasov eq}) is also called 
linearized Boltzmann transport equation without collisional term. 
In Ref. \cite{ota2018} it has been solved numerically 
by preparing the system at equilibrium 
in the presence of a weak stationary potential 
generating a sinusoidal density modulation of a given wavelength.   
Then the potential has been suddenly removed to generate 
a damped time-dependent oscillation 
and hence the speed of sound.

On the contrary, here we directly solve Eq. (\ref{susceptibility}) 
by a fully analytical approach.} 

In Eq. (\ref{susceptibility}) there is a singularity on the 
integration path for $\omega=\mathbf{p}\cdot\mathbf{k}/m$.
In order to attach a meaning to the integral, we must 
interpret $\omega$ as a complex quantity, i.e. 
$\omega = \omega_R + i\omega_I$, where $\omega_I > 0$ in order to 
avoid an exponential growth of the perturbation \cite{landau}.

Eq. \eqref{susceptibility} can be further simplified by assuming, 
without loss of generality, that $\mathbf{k}\parallel \hat{e}_x$, i.e.
$\mathbf{k}=(k,0)$. In this way one finds 
\begin{equation}
1-g_{2D}\int {dp_x\over (2\pi \hbar)} 
\frac{\partial \tilde{f}_0(p_x)}{\partial p_x} 
{1\over \frac{p_x}{m} - c } = 0 \;
\label{susceptibility semplificata}
\end{equation}
where $c=\omega/k$ and $\tilde{f}_0(p_x) = 
\int f_0(p_x,p_y) {dp_y/(2\pi \hbar)}$. 
Clearly, from Eq. (\ref{susceptibility semplificata}) one 
can extract the speed $c$ of sound in our collisionless regime. 
This velocity is, in general, a complex number 
such that $c=\omega/k=c_R+i c_I$ with $c_R=\omega_R/k$ and $c_I=\omega_I/k$. 

In the limit of weakly damped wave, i.e. $c_I \ll c_R$, an elegant formulation 
is provided for the real and imaginary part of $c$ 
\cite{baldovin2016}. In particular, one finds 
two coupled equations for the real part $c_R$ and 
the imaginary part $c_I$ of the speed of sound. 
The equation derived from the real part 
of Eq. (\ref{susceptibility semplificata}) reads  
\begin{equation}
1-g_{2D}\mathcal{P}\int {dp_x\over (2\pi \hbar)} 
\Big[\dfrac{\partial \tilde{f}_0(p_x)/\partial p_x}
{p_x/m - c_R}\Big] 
- \pi c_I \frac{\partial\phi(c)}{\partial c}\bigg|_{c_R}
 = 0 \; , 
\label{real part of omega}
\end{equation}
where we denote $\phi(c) = \frac{m g_{2D}}{(2\pi\hbar)} 
\frac{\partial \tilde{f}_0}{\partial p_x}\Big|_{p_x=mc}$ 
and $\mathcal{P}$ means principal value. 

The equation derived from the imginary part of 
Eq. (\ref{susceptibility semplificata}) is instead given by 
\begin{equation}
c_I =  
\frac{\pi \frac{\partial \tilde{f}_0(p_x)}{\partial p_x}\Big|_{p_x=mc_R}}
{\frac{\partial}{\partial c_R} 
\bigg\lbrace \mathcal{P}\int {dp_x\over (2\pi\hbar)} 
\Big[\dfrac{\partial \tilde{f}_0(p_x)/\partial p_x}
{p_x/m - c_R}\Big] \bigg\rbrace} \; .  
\label{imaginary part of omega}
\end{equation}

\begin{figure}[ht!]
\centering
\includegraphics[width=0.99\columnwidth,clip=]{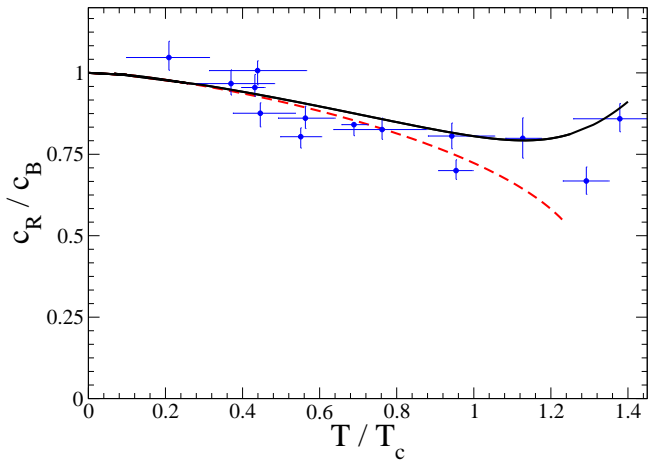}
\caption{Sound velocity $c_R$ in units of $c_B = \sqrt{g_{2D}n/m}$
as a function of the scaled temperature $T/T_c$ 
for $\tilde{g}_{2D} \simeq 0.16$.
The solid black line represents our prediction based on 
Eqs. (\ref{real}) and (\ref{imaginary}) 
while the blue dots are the experimental
data of Ref. \cite{dalibard2017-ictp}. 
The red dashed line is obtained by using 
Eq. (\ref{real}) with $c_I=0$. On the basis of 
universal relations \cite{prokofev2002}, 
for $\tilde{g}_{2D}$ the Berezinskii-Kosterlitz-Thouless 
critical temperature is $T_c = 0.13\;T_B$.}
\label{fig1}
\end{figure}

\textit{Sound velocity for the 2D Bose gas}.
In order to describe the behaviour of the quasi-2D uniform Bose gas 
below or just above the critical temperature, we choose the Bose-Einstein 
distribution function
\begin{equation}
f_0(\mathbf{p}) = \frac{1}{L^2}
\frac{1}{e^{\beta ({p^2\over 2m}+ g_{2D} n - \mu)} -1} \,
\label{bose ansatz}
\end{equation}
as the thermal equilibrium distribution of 2D weakly-interacting 
bosonic atoms with uniform 2D number density $n=N/L^2$, where 
$\beta \equiv (k_B T)^{-1}$, $k_B$ is the Boltzmann constant 
and $T$ is the absolute temperature. Here $\mu$ is the 2D chemical potential 
of the interacting system. Clearly the Hartree interaction 
term $g_{2D} n$ can be formally removed by introducing a shifted 
chemical potential ${\tilde \mu}=\mu - g_{2D}n$.  

The equation of state, relating the shifted chemical potential 
${\tilde \mu}$ to the number density
$n=N/L^2$, is simply derived from the normalization condition
\begin{equation}
N = \int {d^2\mathbf{r} 
d^2\mathbf{p}\over (2\pi \hbar)^2} f_0(\mathbf{p}) \; , 
\end{equation}
resulting in
\begin{equation}
{\tilde \mu} = k_BT \ \ln{\left( 1 -e^{-T_B/T} \right)} 
\label{equation of state}
\end{equation}
where $k_B T_B = 2\pi\hbar^2 n/m$ is the temperature of Bose degeneracy 
and clearly ${\tilde \mu}<0$. 

The analytical computation of the dispersion relation can be simplified 
for temperatures $T\ll T_B$ and, at the same time, 
$c_R^2 \ll k_B T/m$. Within this range of 
parameters one is allowed to write $f_0(\mathbf{p}) L^2 
\simeq k_BT/(p^2/(2m) - {\tilde \mu})$ from which 
$L^2{\tilde f}_0(p_x)=k_BT/(\hbar\sqrt{(p_x^2/m) - 2{\tilde \mu}})$. 
Consequently, the coupled Eqs. \eqref{real part of omega} 
and \eqref{imaginary part of omega} 
for the real and imaginary part of the zero-sound velocity 
respectively read
\begin{widetext}
\begin{equation}
1+ \frac{\tilde{g}_{2D}k_BT}{2\pi} \Big[
\frac{2}{mc_R^2 - 2{\tilde \mu}} 
+ \frac{\sqrt{mc_R^2}}{(mc_R^2-2{\tilde \mu})^{3/2}}
\ln{\bigg(\frac{\sqrt{mc_R^2-2{\tilde \mu}} -\sqrt{mc_R^2}}
{\sqrt{mc_R^2-2{\tilde \mu}} +\sqrt{mc_R^2}}\bigg)} \Big] 
+ \tilde{g}_{2D} k_B T  c_I \frac{mc_R^2  + {\tilde \mu}}
{\sqrt{m}(mc_R^2 -2 {\tilde \mu} )^{5/2}}= 0 \; , 
\label{real}
\end{equation}
\newline
\begin{equation}
c_I = - \frac{\frac{c_R}{\sqrt{m}(mc_R^2 -2{\tilde \mu})^{3/2}}}
{\frac{6\,c_R}{(mc_R^2 -2{\tilde \mu})^2} + 
	\frac{2(mc_R^2 +{\tilde \mu})}{\sqrt{m}(mc_R^2-2{\tilde \mu})^{5/2}}
	\log\Big( 
	\frac{\sqrt{mc_R^2 -2{\tilde \mu}} - \sqrt{mc_R^2}}{\sqrt{mc_R^2 -2
			{\tilde \mu}} + 
		\sqrt{mc_R^2}} \Big)} \; .
\label{imaginary}
\end{equation}
\end{widetext}
By inserting Eq. \eqref{imaginary} in Eq. \eqref{real}
we get an equation for $c_R$.
This equation can be easily solved numerically and, taking into account 
Eq. (\ref{equation of state}), one finds the real part $c_R$ 
of the zero-sound velocity as a function of temperature $T$ and adimensional 
interaction strength ${\tilde g}_{2D}$.  

In Fig. \ref{fig1} we compare the solution of 
Eq. \eqref{real} with the experimental data 
reported in Ref. \cite{dalibard2017-ictp}. 
The agreement between our results and the experimental points is excellent
in the low-temperature regime and still good close to the 
superfluid threshold given by the Berezinskii-Kosterlitz-Thouless 
critical temperature $T_c$. The velocity $c_R$ does not display 
any discontinuity at the critical temperature $T_c$. 
This feature marks a crucial difference with respect to 
first-sound and second-sound velocities calculated within the superfluid 
Landau-Khalatnikov model, { which intrinsically 
relies upon a collisional dynamics of the normal component} 
\cite{ozawa2014,stringari-book}. 
Despite the similar behaviour exhibited far below $T_c$
by the second-sound velocity { $c_2$} \cite{ota2017} 
and our collisionless velocity $c_R$, the former is related 
to the superfluid density and consequently it jumps to zero 
at $T_c$ \cite{ota2017}. 

The dashed line of Fig. \ref{fig1} is obtained by using 
Eq. (\ref{real}) with $c_I=0$. Comparing the dashed line 
with the solid line, which is instead derived solving the coupled 
Eqs. (\ref{real}) and (\ref{imaginary}), one clearly sees 
the increasingly relevant role played by the imaginary part $c_I$ 
(the so-called Landau damping) above $T_c$. 

\begin{figure}[ht!]
\centering
\includegraphics[width=0.99\columnwidth,clip=]{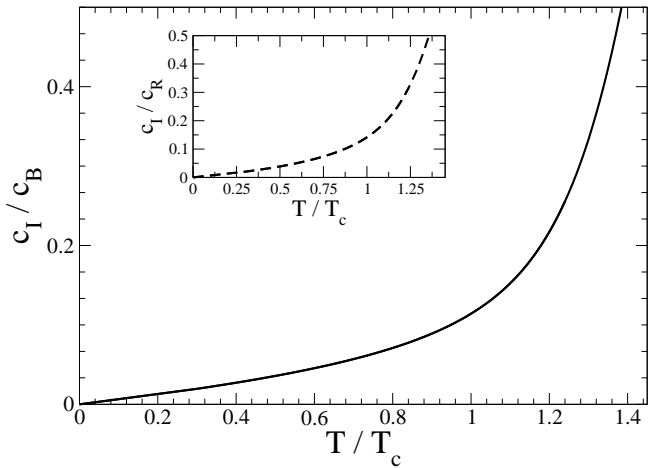}
\caption{\textit{Main panel}: Imaginary 
part $c_I$ of the sound velocity in units of $c_B= \sqrt{g_{2D}n/m}$ 
as a function of the scaled temperature $T/T_c$ 
for $\tilde{g}_{2D} \simeq 0.16$. 
The solid black line is obtained from Eq. \eqref{imaginary} where $c_R$
has derived by solving Eq. \eqref{real}. \textit{Inset}: Ratio 
between the imaginary and the real part of $c$ 
as a function of the temperature.}
\label{fig2}
\end{figure}

In Fig. \ref{fig2} we report the absolute value of $c_I$ as predicted by
Eq. \eqref{imaginary}, where $c_R$ is simply the solution of 
Eq. \eqref{real}. We remark that Eq. \eqref{real part of omega} 
and \eqref{imaginary part of omega} 
are derived by assuming a weakly-damped perturbation, i.e. $c_I \ll c_R$.
>From Fig. \ref{fig2} it appears clear that our approximation scheme is 
surely reliable for low temperatures, where Landau damping 
plays a negligible role, but also in the proximity 
of the transition temperature $T_c$. Quite remarkably, the 
rapid growth of $c_I/c_R$ with the temperature $T$ above $T_c$ 
is in agreement with the large damping of sound oscillations found in 
Ref. \cite{dalibard2017-ictp}. A large value of $c_I/c_R$ also 
signals the breaking of our theoretical scheme, also if the theoretical 
results reproduce the experimental data.

\textit{Conclusions}. 
We have analyzed the sound propagation in collisionless bosonic gases 
assuming a 2D configuration. By solving the linearized 
2D Landau-Vlasov equation in the degenerate 
regime, where bosonic statistical effects play a relevant role, 
we have derived an integral equation for the speed of sound 
as a function of temperature and interaction strength. 
From this integral equation we have obtained two coupled 
algebraic equations for the real and imaginary part of the sound velocity. 
We have then compared our theoretical results with experimental 
data of a recent experiment \cite{dalibard2017-pra,dalibard2017-ictp}, 
where the $^{87}$Rb atoms of the bosonic cloud are expected 
to be in the collisionless regime. This expectation is fully 
confirmed: the agreement betweeen our theory and the experiment 
is very encouraging. { Our theoretical analysis strongly 
suggests that the density perturbation used in the experiment of Ref. 
\cite{dalibard2017-ictp} has excited the "bosonic zero sound", 
i.e. the sound of a collisionless bosonic fluid. 
For a superfluid system, a density perturbation 
can be used to excite the second sound only if the system 
is weakly interacting and collisional \cite{ota2017}. 
By increasing the interaction strength $g_{2D}$ the 2D bosonic system 
enters in the collisional regime where the 
Landau-Vlasov equation (\ref{2d boltzmann-vlasov}) 
loses its validity. The collisional 
regime is in fact correctly described by the 
two-fluid model of Landau-Khalatnikov which reduces to the usual 
hydrodynamics above the critical temperature $T_c$.}  

During the final stage of this work, a theoretical preprint 
on the same topic appeared \cite{ota2018}. 
The conclusions of Ref. \cite{ota2018}, based on stochastic 
Gross-Pitaevskii equation and dynamic response function, 
are similar to ours. 

{\it Acknowledgments}. 
The authors thank Franco Dalfovo for useful discussions. 
LS acknowledges for partial support the FFABR grant of 
Italian Ministry of Education, University and Research.


\begin{thebibliography}{99}

\bibitem{kirkpatrick1983} T. R. Kirkpatrick and J. R. Dorfman,
Phys. Rev. A \textbf{28}, 2576(R) (1983).

\bibitem{kirkpatrick1985-lowtemp} T. R. Kirkpatrick and J. R. Dorfman,
J. Low Temp. Phys. \textbf{58}, 301-332 (1985).
	
\bibitem{zaremba1998} E. Zaremba, A. Griffin and T. Nikuni, 
Phys. Rev. A \textbf{57}, 4695 (1998).

\bibitem{nikuni1999} T. Nikuni, E. Zaremba, and A. Griffin, 
Phys. Rev. Lett. \textbf{83}, 10 (1999).

\bibitem{zaremba1999-lowtemp} E. Zaremba, T. Nikuni, and A. Griffin,
J. Low Temp. Phys. \textbf{116}, 277-345 (1999).

\bibitem{griffin-book} A. Griffin, T. Nikuni, and E. Zaremba,
\textit{Bose-Condensed Gases at Finite Temperatures} 
(Cambridge University Press, New York, 2009).

\bibitem{vlasov1945} A. Vlasov, J. Phys. USSR \textbf{9}, 25 (1945).

\bibitem{landau1946} L. D. Landau, J. Phys. USSR \textbf{10}, 25 (1946)

\bibitem{landau} L. D. Landau and E. M. Lifshitz,
\textit{Course of Theoretical Physics 10: Physical Kinetics}
(Pergamon International Library, Exeter, 1981). 

\bibitem{baym-book} L. Kadanoff and G. Baym, 
\textit{Quantum Statistical Mechanics: 
Green's Function Methods in Equilibrium and Nonequilibrium Problems}
(W.A. Benjamin, New York, 1962).

\bibitem{mermin} N. D. Mermin and H. Wagner, 
Phys. Rev. Lett. {\bf 17}, 133 (1966).

\bibitem{hohenberg} P.C. Hohenberg, Phys. Rev. {\bf 158}, 383 (1967).

\bibitem{berezinskii} V. L. Berezinskii, Sov. Phys. JETP {\bf 34}, 
610 (1972).

\bibitem{kosterlitz} J. M. Kosterlitz and D. J. Thouless, 
J. Phys. C: Solid State Phys. {\bf 6}, 1181 (1973).

\bibitem{dalibard2017-pra} V. P. Singh, C. Weitenberg, J. Dalibard, and
L. Mathey, Phys. Rev. A \textbf{95}, 043631 (2017).

\bibitem{dalibard2017-ictp} J.L. Ville, R. Saint-Jalm, 
E. Le Cerf, M. Aidelsburger, S. Nascimb\'ene, J. Dalibard, and J. Beugnon, 
arXiv:1804.04037 (2018).

\bibitem{ota2017} M. Ota and S. Stringari, Phys. Rev. A \textbf{97}, 
033604 (2018).

\bibitem{prokofev2001} N. Prokof'ev, O. Ruebenacker and B. Svistunov,
Phys. Rev. Lett. \textbf{87}, 270402 (2001).

\bibitem{prokofev2002} N. Prokof'ev, O. Ruebenacker, and B. Svistunov,
Phys. Rev. A \textbf{66}, 043608 (2002).

\bibitem{ota2018} M. Ota, F. Larcher, F. Dalfovo, L. Pitaevskii, 
N.P. Proukakis, and S. Stringari, arXiv:1804.04032 (2018).

\bibitem{baldovin2016} F. Baldovin, A. Cappellaro, E. Orlandini, 
and L. Salasnich, J. Stat. Mech. 063303 (2016).

\bibitem{ozawa2014} T. Ozawa and S. Stringari, 
Phys. Rev. Lett. \textbf{112}, 025302 (2014).

\bibitem{stringari-book} L. P. Pitaevskii and S. Stringari, 
\textit{Bose-Einstein Condensation and Superfluidity}
(Oxford University Press, Oxford, 2016).

\end{thebibliography}
\end{document}